\documentclass[paper,12pt]{JHEP}
\usepackage[centertags]{amsmath}
\usepackage{amsfonts} \usepackage{amssymb} \usepackage{amsthm}
\usepackage{bbm}
\usepackage{bm}

\newcommand{\be}{\begin{equation}}
\newcommand{\ee}{\end{equation}}
\newcommand{\bea}{\begin{eqnarray}}
\newcommand{\eea}{\end{eqnarray}}
\newcommand{\ba}{\begin{array}}
\newcommand{\ea}{\end{array}}
\def\nn{\nonumber}

\newcommand{\ft}[2]{{\textstyle\frac{#1}{#2}}}
\newcommand{\Z}{{\mathbb Z}}
\newcommand{\R}{{\mathbb R}}

\title{Exotic prepotentials from D(-1)D7 dynamics}
\author{\parbox{11.5cm}{ Francesco Fucito$^1$ Jose F. Morales$^1$ and  Rubik Poghossian$^{1,2}$}\\
\vspace{0.3cm}

1.  I.N.F.N.,  Universit\`a di Roma Tor Vergata\\ 
Via della Ricerca Scientifica, I-00133 Roma, Italy\\
2. Yerevan Physics Institute\\Alikhanian Br. 2\\
0036 Yerevan, Armenia\\
\email{fucito, morales, poghosyan@roma2.infn.it}
}
\abstract{
We compute the partition functions of D(-1)D7 systems describing the
 multi-instanton dynamics of $SO(N)$ gauge theories in eight dimensions.  
 This is the simplest instance of the so called exotic instantons.
  In analogy with the Seiberg-Witten theory in four
space-time dimensions, the prepotential and correlators in the chiral ring are derived 
via localization formulas and found to satisfy relations of the Matone type. Exotic prepotentials of SO(N) 
gauge theories with ${\cal N}=2$ supersymmetries in four-dimensions are also discussed. 
}
\begin{document}
\section{Introduction}
Extracting the standard model or some supersymmetric (SUSY) extension of it from a D-brane construction
has been the focus of much recent work (see \cite{Blumenhagen:2005mu,Blumenhagen:2006ci,Marchesano:2007de} for recent reviews).
In this framework,
it has been recently understood that certain non perturbative effects may lead to interesting new phenomena
in low energy theories like a generation of a Majorana mass term 
or of Yukawa couplings \cite{Blumenhagen:2006xt,Ibanez:2006da},  not to mention new possible patterns of SUSY breaking
\cite{Berenstein:2005xa}\nocite{Franco:2005zu,Bertolini:2005di,Diaconescu:2005pc,Heckman:2008es,Aharony:2007db,Buican:2008qe,Cvetic:2008mh}-\cite{Bianchi:2009bg}.
This proposal was further sharpened by the observation that to really have such effects, orientifold planes 
\cite{Argurio:2007qk}\nocite{Argurio:2007vqa,Bianchi:2007wy,Blumenhagen:2007zk,Ibanez:2007tu}-\cite{Ibanez:2007rs} or 
closed string fluxes \cite{Billo':2008sp,Billo':2008pg}
need to be introduced.

To the extent of understanding non-perturbative effects in systems of D-branes a derivation of four dimensional instantons
in terms of D-branes was carried out in \cite{Witten:1995im,Douglas:1995bn,Douglas:1996uz}. These results where re-discussed
in \cite{Billo:2002hm} where the field theory results   
were obtained by string methods  and a careful parallel between
the ADHM formalism and the
system formed by bound states of parallel D(-1) and D3 branes was carried out. The reader should be aware that the non perturbative
effects alluded to in the
previous paragraph are not of this type. In general bound states of intersecting branes at angles (with more than four Neuman-Dirichlet
directions) or branes hosting generic 
non parallel magnetic fluxes  lack those bosonic moduli related to the instanton sizes and gauge orientations. This type of instantons
have been called exotic and  besides their phenomenological applications are interesting in itself since they are relevant for many
non perturbative effects in string theory. 
 
We will then focus our attention on the D(-1)D7 system in the presence of an orientifold O7 plane. 
This is the simplest supersymmetric instance of an exotic instanton. We choose the orientifold projection in such a way to
get an $SO(N)$ gauge theory with $SO(k)$ exotic instantons which carry the right number of zero modes to generate a 
potential\cite{Argurio:2007vqa,Bianchi:2007wy}.
From the low energy gauge theory viewpoint, this is an eight dimensional instanton. 
As it is well-known, extending the idea of self-duality to dimensions greater than
four is far from obvious. Self-duality in four dimensions is tantamount to say that the field strength is a $(1,1)$ complex form of
Einstein-Khaeler type. This was the starting
point of the first explorations in this field \cite{Corrigan:1982th}. Later a solution to the quadratic Yang-Mills (YM) action
in eight dimensions was found \cite{Fubini:1985jm} with gauge group $SO(7)$. Finally an $SO(8)$ gauge connection was
exhibited \cite{Grossman:1989bb,Grossman:1984pi} which was the generalization of the Hopf map $S^7\xrightarrow{S^3} S^4$ 
in four dimensions  
to eight dimensions where the $SO(8)$ gauge bundle  is thought as the  Hopf map $S^{15}\xrightarrow{S^7} S^8$.
This latter solution has been recently related to D(-1) instantons on the SO(8) gauge theory living in a D7 O7 worldvolume  \cite{Billo:2009gc}:
this interpretation is  supported by the observation \cite{Billo:2009gc}
   that in the limit of instanton zero size the quadratic YM term vanishes. Moreover the 
   quartic term, $F^4$, becomes proportional to the fourth Chern class of the
 gauge bundle thus matching the D(-1) action.  
    This proposal have been  put on solid grounds in \cite{Billo':2009di} where 
 instanton corrections to $F^4$-terms were computed in complete analogy with the four dimensional case
  \cite{Moore:1998et}\nocite{Nekrasov:2002qd,Flume:2002az,Bruzzo:2002xf,Losev:2003py,Nekrasov:2003rj}-\cite{Flume:2004rp} 
finding agreement with  the heterotic results  
\cite{Lerche:1988zy}\nocite{Lerche:1987qk,Lerche:1998nx,Bachas:1997mc,Bachas:1996bp,Forger:1998ub,Gava:1999ky}-\cite{Gutperle:1999xu}.
We remark that an ADHM construction of the moduli space of these "exotic"  eight dimensional
   instantons is missing and therefore these D-brane techniques are at present the only way to investigate this physics.
  
 The $SO(8)$ gauge theory in eight dimensions  is very special. The theory is conformal, in the
 sense that the coupling $\tau_4$ of the $F^4$-term does not run. Conformal invariance implies  
 that $\tau_4$ is constant over the moduli space since it cannot depend on dimensionful quantities such as the
 vevs of the scalar field. This is very different to what one expect in the non conformal case with gauge group $SO(N)$, $N>8$,
 where the instanton measure acquires a dimension and the $F^4$ coupling becomes a non trivial function of the
 casimirs parametrizing the moduli space. The aim of this paper is to address the study of multi-instanton corrections to the
 effective action  of such gauge theories. We will also consider  the case of  ${\cal N}=2$ gauge theories
 in four dimensions which arises from placing the D(-1)D7 system at a $\R^4/\Z_2$ singularity that freezes the dynamics
 along the orbifold four-plane. How to test these results against heterotic computations with non trivial
Wilson lines remains an open and challenging task\footnote{We thank M. Bianchi for discussions on this issue.}.
 
This is the plan of the paper: in Section 2 and Section 3 we review the main features of the eight dimensional instanton 
and the  localization algorithm. In Section 4 and Section 5 we compute the exotic prepotentials 
 describing the dynamics of  $SO(N)$ gauge theories in $d=8$ and $d=4$ dimensions respectively. 
  Finally in Section 6 we compute the correlators of the chiral ring.

\section{Eight dimensional instantons }

 In this section we review the proposal in \cite{Billo:2009gc,Billo':2009di} for a D(-1)D7 description of the $SO(8)$ 
 eight-dimensional instantons  and extend it to  the non conformal case with gauge group SO(N).
 The field content of maximal supersymmetric YM theory in eight dimensions includes a gauge boson with
 field strength $F$, a complex scalar $\phi$ and two fermions of opposite chirality. 
 In complete analogy with the ${\cal N}=2$ SYM in
 $d=4$ dimensions the chiral dynamics of the SYM theory in eight dimensions
 is described by a prepotential ${\cal F}(\Phi)$ in terms of which the effective action can be written as
  \bea
 S_{\rm eff} &=& \int d^8 x d^8 \theta\,  {\cal F}(\Phi)  +{\rm h.c.}
  \eea
 with
 \be
  \Phi=\phi+\sqrt2\theta \Psi+F_{\mu\nu} \theta \gamma^{\mu\nu} \theta+\ldots
 \ee
  the chiral eight-dimensional superfield.  At the classical level ${\cal F}_{\rm cl}= i\tau_4 \, {\rm Tr}\, \Phi^4$
  and the effective action becomes
 \be
 S_{\rm eff}=  {\rm Re} \,\tau_4\, \int_{\R^8}  d^8 x \, t_8\, F^4+{ i  {\rm Im} \,\tau_4\,  }
  \int_{\R^8}  F\wedge F\wedge F \wedge F
\ee
with
\be
\tau_4={\theta\over 2\pi}+ \,i{4! (2\pi)^3\over g^2}
 \ee
  and where $t_8$ is the invariant eight-rank tensor of SO(8). As we said in the introduction, D(-1) instantons
  are identified with zero size instantons with vanishing quadratic YM action and therefore we will always discard this term
  from the effective action.      The gauge theory can be realized in terms of a stack of N D7-branes on
  top of a O7-plane.   Instantons correct the prepotential function ${\cal F}(\Phi)$ and then the effective action.
   Instantons in the eight-dimensional gauge theory are realized in terms of D(-1) branes
  with open strings describing  the instanton moduli space and the D(-1)D7 action gives the gauge dynamics around the instanton
background. By instantons here  we refer to
solutions  of the Yang-Mills equations with action
  \be
   S_{\rm cl}=2 \pi\, \tau_4 \,  k
       \ee
    with $k$ the fourth Chern class
    \be
    k={ 1\over 4!(2\pi)^4}
  \int_{\R^8}  {\rm Tr}\, F\wedge F\wedge F \wedge F
    \ee
 An explicit solution in this class can be written as\footnote{ Using the properties of SO(8) gamma functions one finds that $F$
 satisfy   $*(F\wedge F)=F\wedge F $ and   $*F=T\wedge F$ with
 $ T_{\mu\nu\sigma\rho}= \eta\, \gamma_{[\mu} \,\gamma_{\nu}^\dagger\,\gamma_\sigma \, \gamma_{\rho]}^\dagger\, \eta$
with $\eta$ a fixed eight dimensional spinor.  These two relations define possible generalizations of the concept
of four-dimensional self-duality to eight dimensions.   }
 \be
F_{\mu \nu}=-{ 2 \rho^2\over (x^2+\rho^2)^2} \,\,   \gamma_{\mu\nu}
\ee
 with
\be
\gamma_{\mu\nu}=\Upsilon
\left(\begin{array}{cc}
0 & 0\\
0 & \gamma_{\mu\nu}^{SO(8)}
\end{array}\right) 
\Upsilon^T
 \label{gammam}
\ee
and $\Upsilon\in SO(N)/SO(N-8)$ parametrizing the orientation of the SO(8) instanton inside SO(N).
$\gamma_{\mu\nu}^{SO(8)}=\gamma_{[\mu}\gamma_{\nu]}^\dagger$ with $\gamma_\mu$ the $SO(8)$ gamma matrices
satisfying $\gamma_{(\mu}\gamma_{\nu)}^\dagger=\delta_{\mu\nu}$.
   This is the solution originally found in \cite{Grossman:1984pi,Grossman:1989bb}. In the limit $\alpha^\prime, \rho\to 0$ the quadratic YM action
   evaluated at this solution vanishes and the quartic term matches that of the D(-1) instanton with $\theta=C_0$ the RR 0-form and
   $g^2=g_s$ the string coupling \cite{Billo:2009gc}.   

    The study of the D(-1)D7 dynamics follows closely that of its four-dimensional D(-1)D3 analog with some important differences.
    First the O7-orientifold projection acts with the same sign on the D7 and D(-1) Chan-Paton indices. 
    This implies in particular that the symmetry group of k D(-1) instantons in the $SO(N)$ gauge theory coming from the D7-branes is $SO(k)$.
    This is in contrast with the four-dimensional case where $SO(N)$ instantons carry an $Sp(k)$ symmetry group.
      Secondly, unlike in the D(-1)D3 case, open strings between D(-1) and D7 branes have no bosonic zero modes and therefore interactions
    between the two brane stacks are mediated only via the fermionic field $\mu$  (the Ramond ground state) 
    in the bifundamental of the $SO(N)\times SO(k)$ symmetry group. 
      Like in the four-dimensional case, the correlators in the gauge theory can be written as the moduli space
   integral \cite{Bianchi:2007wy,Billo':2009di}
      \be
      \langle {\cal O} \rangle = {1 \over Z} \sum_k \, \hat q^k \, \ell_s^{-k\beta_4} \int d\mathfrak M_{k,N}\, e^{-S_{k}-\mu^T \phi \mu}\, {\cal O}
      \ee
    with $\ell_s=\sqrt{\alpha'}$ the typical string length, $\hat q=e^{2\pi i \tau_4}$ with $\tau_4$ the $F^4$-coupling and $\beta_4$ its one-loop 
beta function.  
    $S_{k}$ is the instanton action following from dimensional reduction of ${\cal N}=1$ $d=10$ SYM down to zero dimensions
   and $ \mathfrak M_{k,N}$ the D(-1)D7 moduli space. The term  $\ell_s^{-k\beta_4}$ comes from the one-loop vacuum amplitudes 
   and it compensates the dimension of the instanton measure. 
   Finally $Z=\langle \mathbb{I} \rangle $ is the instanton partition function. Later we will also consider correlators of the form
    $\langle {\rm Tr} \phi^J  \rangle$. In particular the basic correlator  $\langle {\rm Tr} \phi^4  \rangle$ will be related to the
    prepotential of the eight dimensional theory. 
    
We remark that $\Phi$ couples to the D(-1) instantons only through the Yukawa coupling $\mu^T \phi \mu$. Integration
over $\mu$ leads to a polynomial dependence of $\phi$. 
 This is in sharp contrast with the standard gauge
 instanton potentials which fall off to zero in the limit of large vev of the scalar field.   Another important difference
 with the four-dimensional case is that the weak coupling  regime is defined by the limit $\ell_s\to 0$ rather than $\Lambda\to 0$.
 This implies in particular that a weak coupling analysis is reliable only in the case $\beta_4 \leq 0$. This is the case for
 the $SO(N)$ gauge theories with maximal supersymmetry and $N\geq 8$ where the beta function $\beta_4=(8-N)$ is
 negative (see \cite{Bianchi:2000vb} for computations of the $F^4$ beta function for an arbitrary field content).

    \section{Localization formulae}

In this Section we review the localization algorithm developed in \cite{Moore:1998et,Nekrasov:2002qd,Flume:2002az,Bruzzo:2002xf}
to  study integrals over  the instanton moduli space. The main idea
is the identification of an equivariantly deformed BRST operator $Q_\xi$, satisfying $Q_\xi^2={\cal L}_\xi$
with ${\cal L}_\xi$ a Lie derivative on the moduli space along a vector field $\xi$ belonging to the Cartan of the
$SO(K)\times SO(N)\times SO(8)$ symmetry group. $\xi$ can be parameterized by $\xi=(\chi_i,a_u,\epsilon_\ell)$ with
$i=1,..k$, $u=1,..n$, $\ell=1,..4$ $(n=[N/2], k=[K/2])$.  Choosing a generic $\xi$ the symmetry group is
broken to its Cartan part and the
integral is given by the contributions at isolated critical points (the poles in a contour integral over $\chi_i$).
Physical quantities in the gauge theory are defined by taking the limit $\epsilon_\ell\to 0$ in order to eliminate
the singularity arising from the D(-1) branes all superposed at the origin.
More precisely,  the prepotential $F(\Phi)$ is given by the formula 
\be F(a_u) =\lim_{\epsilon_\ell \to 0} \, \left[ \epsilon_1 \epsilon_2\epsilon_3\epsilon_4\, \ln
\sum_k Z_k (a_u,\epsilon_\ell) q^k\right]  \ee
with $Z_k$ the instanton partition function. 
 Keeping $\epsilon_\ell$ finite one finds the gravitational corrections to the Yang-Mills action \cite{Billo:2006jm}.
  The instanton partition function $Z_k(a_u,\epsilon_\ell)$  can be written as an integral over the instanton moduli space.
 The integral can be performed using the localization formula 
  \be
Z_k=\int d\mathfrak{M}_{k,N} e^{-S_{\rm inst} }=\int  {d^k \chi_i \over {\rm Sdet} Q_\xi^2 } =\int d^k \chi_i  
\prod_{\Phi} \lambda_{\Phi}(\chi,a,\epsilon)^{(-)^{F_{\Phi}+1}}
\ee
 where   $\Phi$ labels the Q-multiplet pairs $(\Phi,\Psi)$  related by the BRST transformations
 \be
 Q \, \Phi=\Psi \qquad Q\Psi=\lambda_{\Phi}(\chi,a,\epsilon) \, \Phi
 \ee
  $F_{\Phi}=0\,(1)$ for $\Phi$ a bosonic (fermionic) field
  and $\lambda_\Phi$ is the eigenvalues
 of $Q^2_\xi$.  It is important to notice that using $Q\sim \ell_s^{-{1\over 2}}$ the dimension of the measure of the moduli space can be written as
  \be
 \left[ d\mathfrak{M}_{k,N}\right]  =\ell_s^{{1\over 2}(n_F-n_B)}
 \ee
  with $n_B,n_F$ the number of Q-multiplets with lowest component a boson and a fermion respectively.

The topological theory for the system discussed in the previous section,
describes the excitations of open strings connecting the various branes with at least one end on the D(-1) instanton.
 We denote the fields by $\Phi_{\cal AB}$ where the index ${\cal A}=(I,U)$ runs over all possible boundaries $I=1,....K$ (number
 of D(-1) branes), $U=1,...N$ (number of D7-branes).
In presence of an O7 plane, let us say at $x=0$ in the transverse plane, the branes are distributed symmetrically
with respect to it. We denote by $x_{\cal A}=(\chi_I,a_U)$ the positions of the various branes
\bea
\chi_I &=& (\chi_1,....,\chi_k;-\chi_1,...,-\chi_k; 0)\nn\\
a_U &=& (a_1,....a_n,-a_1,....,-a_n;0)
 \label{positions}
\eea
 The last "0" should be omitted in the case of an  even number of branes. The BRST operator is defined by
 equivariantly deforming the SUSY algebra by the $SO(K)\times SO(N) $
 brane symmetry. In addition complete localization requires also a $U(1)^3$ deformation  
 inside the Lorentz $SO(8)$ group parametrized by $\epsilon_\ell$, $\ell=1,\ldots 4$ with $\sum_\ell \epsilon_\ell=0$.
     More precisely the $Q^2$-eigenvalue of a field
  $\Phi_{\cal AB}$ can be written as
  \be
  \lambda_{\Phi}=x_{\cal A}-x_{\cal B}+q_{\Phi}
  \ee
  with $q_{\Phi}$ the $U(1)^3$ charge  of the given field.
    Taking into account that the presence of the orientifold halves the degrees of freedom in the covering space
  the  partition function can then be written as
  \be
  Z_K=\int d^k{\chi_i} \, \prod^\prime_{{\cal A,B}, \Phi}  \,   (x_{\cal A}-x_{\cal B}+q_{\Phi})^{{1\over 2}(-)^{F_\Phi+1} }
  \prod_{{\cal A}, \Phi}  \,   (2\, x_{\cal A}+q_{\Phi})^{{1\over 2}  (-)^{F_\Phi+1}\delta_{\Phi} } \label{zgeneral}
  \ee
   with $\delta_{\Phi}=\pm$ depending on whether the field is even or odd under the orientifold projection. The
   primed product runs  over  all ${\cal A,B}$ pairs with at least one index  on the D(-1) instantons.
    The second contribution comes from open strings connecting the D-brane ${\cal A}$
   to its image.    It is important to notice that despite the explicit appearance of square roots,
   from (\ref{positions})  one can see that each eigenvalue appears twice and therefore
   the final answer contains no square roots.

    The BRST transformations for the various strings under consideration are:

    \begin{itemize}

    \item{D(-1)D(-1) open strings}

    \bea
    Q\, B_{\ell; IJ} &=& M_{\ell; IJ}  \qquad Q \,M_{\ell; IJ} =(\chi_{IJ}+ \epsilon_\ell )\, B_{\ell; IJ}\nn\\
    Q \,\lambda_{c; IJ} &=& D_{c; IJ}  \qquad Q\, D_{c; IJ} =(\chi_{IJ}+  s_c ) \,\lambda_{c; IJ}
   \eea
    with $\ell=1,..4$, , $c=1,..4$ and
    \bea
s_1=\epsilon_2+\epsilon_3 \quad\quad
s_2=\epsilon_1+\epsilon_3 \quad\quad
s_3=\epsilon_1+\epsilon_2\qquad   s_4=0\nn\\
\epsilon_1+\epsilon_2+\epsilon_3+\epsilon_4=0
\eea
  In writing the BRST multiplets we group the seven ADHM constraints $D_s$, $s=1,...7$ and the
component $\bar\chi$ into four complex fields denoted by $D_c$. In doing this we should remind that the
 zero eigenvalues associated to the diagonal field components  $(\lambda_4,D_4)_{II}$ should be omitted from
 the determinant. Alternatively the contribution of this pair to the partition function can be thought as coming from
 the Vandermonde determinant resulting from bringing the field $\chi$ into its diagonal form.

   The orientifold projection project  $a_\ell$ and $\lambda_c$   on symmetric and antisymmetric matrices respectively i.e.
    \be
    \delta_{a}=+  \qquad \delta_{\lambda}=-
    \ee
    This is consistent with the fact that the ADHM contraints $D_c$ can be written in terms of  commutators of $a_\ell$.

    \item{D(-1)D7 open strings}

 \bea
    Q\, \mu_{I U} &=&  h_{I U}  \qquad Q \, h_{I U}    =(\chi_{I}-a_U)
    \, h_{I U}
   \eea
 The field $h$ is an auxiliary field needed to close the Q-algebra \cite{Fucito:2001ha}.
\end{itemize}

\section{D(-1)D7 on $\R^{10}$ }

In this Section we consider a system of $K$ D(-1) branes, $N=2n$ D7-branes and an O7-plane
realizing a maximal SUSY  $SO(2n)$ gauge theory in eight dimensions.  
The total symmetry group is then $SO(2n)\times SO(K)$. Fields like the fermionic ADHM auxiliary fields $\lambda_c$ 
 transforming in the adjoint of SO(K) group are described  by antisymmetric
matrices while instanton positions $B_\ell$ are given in terms of symmetric matrices.
  The field content in the instanton moduli space is summarized in the following table:
  \begin{equation*}
\begin{array}{|c|c|c|c|c|}
\hline
 (\Phi,\Psi)  & (-)^{F_{\Phi}} & \delta_\Phi    &\mbox{multiplicity} & q_\Phi\\
\hline \hline
  (B_\ell,M_\ell ) & + & + &8\,\ft12\,K(K+1) & \epsilon_\ell \\
  \phantom{\vdots}(\lambda_c ,D_c) & - & - &8\,\ft12\,K(K-1 ) & s_c\\
  (\phantom{\vdots}\mu,h) & -& 0 & 2 n K & 0 \\
\hline
\end{array}
\end{equation*}
Plugging these data into the general formula (\ref{zgeneral}) one finds the
partition function
 \begin{eqnarray}
Z_{K} &=& {\cal N}_K\, \int \prod_{i=1}^{k} \frac{d \chi_i}{2\pi i} \,
\prod_{I,J}^K \left[ {P(\chi_{IJ}  ) \over Q(\chi_{IJ} )}
\right]^{1\over 2}  \prod_{I=1}^{K} \, \left[{M(\chi_I ) \over
 P(2\chi_I )Q(2\chi_I )}\right] ^{1\over 2}
\label{partzk} \eea
 with $\chi_{IJ} =\chi_I - \chi_J$ and
\begin{eqnarray}
P(x)  &=&   x^{1-\delta_{x,0}}  \prod_{a=1}^3(x+s_a),\nn\\
 Q(x)&=&
\prod_{\ell=1}^4(x+\epsilon_\ell)\nonumber\\
 M(x)&=&   \prod_{u=1}^n (x+a_u)
\end{eqnarray}
 These polynomials give the contribution of the fields $\lambda_c$, $B_\ell$ and  $\mu$ respectively.
Notice that $(-)^{F_{\Phi}}=\delta_\Phi$ for  $\Phi=\lambda_c,B_\ell$ explaining why both contributions $P(2\chi_I),Q(2\chi_I)$ come in the
denominator.

  Setting $K=2k$ and $K=2k+1$ and using (\ref{positions}) one finds
 \bea
Z_{2k} &=& {\cal N}_{2k}\, \int \prod_{i=1}^{k} \frac{d \chi_i}{2\pi i}
\,   \prod_{i\leq j}^k {P_2(\chi^-_{ij}  ) P_2(\chi^+_{ij}  ) \over
Q_2(\chi^-_{ij} ) Q_2 (\chi^+_{ij} ) }   \prod_{i=1}^{k} \,  {M_2(\chi_i)
\over
 P_2(2\chi_i)    }  \label{zz}    \\
 Z_{2k+1} &=&  {\cal N}_{2k+1}\, {M(0) \over
    Q_2(0) }   \int \prod_{i=1}^{k}
\frac{d \chi_i}{2\pi i} \,   \prod_{i\leq j}^k {P_2(\chi^-_{ij}  ) P_2(\chi^+_{ij}  ) \over
Q_2(\chi^-_{ij} ) Q_2 (\chi^+_{ij} ) }   \prod_{i=1}^{k} \,  {M_2(\chi_i)   P_2(\chi_i )      \over
 P_2(2\chi_i)    Q_2(\chi_i)}       \nn
\end{eqnarray}
with
\begin{eqnarray}
P_2(x)  &=&   (-x^2)^{1-\delta_{x,0}}  \prod_{a=1}^3(s^2_a-x^2),\nn\\
 Q_2(x)&=&
\prod_{\ell=1}^4(\epsilon_\ell^2-x^2)\,,\nonumber\\
 M_2(x)&=&   \prod_{u=1}^n (a_u^2-x^2)\,,\nn\\
 {\cal N}_{2k}&=&{2^{4 (2k)}\over 2^k k!}\,\,;\quad {\cal N}_{2k+1}={2^{4(2k+1)}\over 2^kk!}\,.
\end{eqnarray}
Integrals over $\chi_i$ should be supplemented with a pole prescription. Following the four dimensional analogy we take
$\epsilon_\ell \to \epsilon_\ell+i\delta_\ell$ with $\delta_1>> \delta_2>>\delta_3>>\delta_4$.   
The prepotential of the eight dimensional theory can be extracted
from the relation 
  \be F(a_u,\epsilon_\ell) =\epsilon_1 \epsilon_2\epsilon_3\epsilon_4\, \ln
\sum_K Z_K (a_u,\epsilon_\ell) q^K 
\ee 
An explicit evaluation of the integrals for the first few instanton contributions leads to
 \bea
&& F(a,\epsilon)= \nn \\ && 8 \sqrt{ A_n}  ( q+\ft{4}{3} \,q^3 \, A_{n-4}-\ft56\, q^3\,
A_{n-5} g_2 + \ft{1}{96}\, q^3\,A_{n-6} (25 g_2^2+34g_4)+\ldots )\nn\\
~~~&& +q^2(  -2 A_{n-2}+\ft14\, A_{n-3} g_2-\ft{1}{64}  A_{n-4}
(g_2^2+2 g_4)   +\ldots )
\label{prepon} \eea
with $A_{m}$, $m=1,...n$  the
$m$th elementary symmetric functions of the variables
$a_1^2,\ldots ,a_n^2$:
\bea
A_s &=&\sum_{i_1< i_2\ldots < i_s}   a_{i_1}^2\ldots  a_{i_s}^2  \nn\\
 A_n &=& a_1^2 \ldots a_n^2    \quad\quad A_0=1 \quad\quad A_{s< 0}=0
\eea
Notice that $A_m$ form a basis for the Casimirs  of $SO(2n)$ .
 Similarly we parametrize the SO(8) Casimirs in terms of
    \bea
g_{2m}=\sum_{\ell=1}^4\epsilon_\ell^{2m}
\eea
      In the appendix we present also $4$ and $5$ instanton contributions.

 The effective action follows by replacing  in the prepotential $F$ the lowest components
$a_u$, $\epsilon_\ell $ by the corresponding chiral and
gravitational superfields \bea
&& a_u \to \Phi_u=\phi_u+F^u_{\mu\nu} \theta \gamma^{\mu\nu} \theta+\ldots\nn\\
&& \epsilon_\ell \to W_\ell=G_\ell+R^\ell_{\mu\nu} \theta \gamma^{\mu\nu} \theta+\ldots
\eea
 Here we denote by $G_\ell$ the graviphoton with $\ell=1,..4$ and $u=1,..n$ running over the Cartan subgroup
components of the Lorentz $SO(8)$ and gauge $SO(2n)$ symmetry groups respectively .

The prepotential $F(\Phi,W)$ encodes the chiral dynamics of the $d=8$ gauge theory coupled to gravity.
The eight-dimensional effective action follows from $F(\Phi,W)$ upon integration over the chiral superspace variables
  \bea
 S_{\rm eff} &=& \int d^8 x d^8 \theta\,  F(\Phi,W)
 \eea
  In absence of gravity $W=0$, it gives a direct analog of the Seiberg-Witten prepotential for ${\cal N}=2$ gauge theories
in $d=4$ dimensions. For instance the $F^4$ coupling
    \bea
 S_{\rm eff} &=&  \tau_4(a)\, \int d^8 x \,{\rm tr}\, t_8 \,F^4
 \eea
 can be expressed as the fourth derivative of the prepotential
  \be
  \tau_4(A)=\left[\sum_{u=1}^n {\partial^4\over \partial a_n^4} -\frac{1}{12} \left(\sum_{u=1}^n {\partial^2\over \partial a_n^2}\right)^2  \right] F(a,0)
 \ee
    Notice that the dimension of the moduli space is   $L^{-{{\rm dim} \mathfrak{M} \over 2} }$ with
   \be
  \ft12{\rm dim} \mathfrak{M}= \ft12(n_a-n_\lambda-n_\mu)=K(4- n)
  \ee
   This implies that a dimensionless partititon function $Z=\sum_K Z_K q^K$ can be defined taking
  $q=\ell_s^{4-n} e^{2\pi i \tau_4}$.
  The case $n=4$ is special in the sense that q is dimensionless and
the theory is conformal: this is the case that has been studied in details in \cite{Billo':2009di}. We recover
in Appendix  \ref{aso8} their results.

  \section{D(-1)D7 on $\R^6\times \R^4/\Z_2$}
 \label{sr4z2}

The analysis in the last section can be  extended to theories with less supercharges and
lower dimensions. Here we consider the ${\cal N}=2$ case in four-dimensions. This theory can be
realized by considering a set of $2n$ fractional D7-branes wrapping a  $\R^4/\Z_2$-singularity. We choose the
$\Z_2$ orbifold  group to act trivially in the Chan Paton indices. This projects out the gauge components
along the four directions acted by the $\Z_2$
leaving an effective four-dimensional theory with ${\cal N}=2$ supersymmetries.
On the instanton moduli  space the orbifold groups acts like 
   \bea 
    a_{3,4}\to
-a_{3,4}  \qquad   M_{3,4}\to -M_{3,4}   \qquad \lambda_{1,2}\to
-\lambda_{1,2}
  \qquad D_{1,2}\to -D_{1,2}
\eea
 with no action on the Chan-Paton indices and all the other fields invariant.
   In the language of fractional branes
 this corresponds to take  $(n_0,n_1)=(2n,0)$ D7-branes and $(k_0,k_1)=(K,0)$
 D(-1) branes.

 The partition function follows now
from the previous results by simply suppressing the contribution of the odd fields.
The results are given again by (\ref{zz})  but
with the characteristic functions replaced by
 \begin{eqnarray}
 P_2(x)  =   x^{2-2\delta_{x,0}} (x^2- \epsilon^2)\nn\\
    Q_2(x)= (x^2-\epsilon^2_1) (x^2-\epsilon^2_2)  \nonumber\\
  M_2(x)=   \prod_{u=1}^n (x^2-a^2_u)
\end{eqnarray}
and $\epsilon=\epsilon_1+\epsilon_2$.

Now the prepotential defines the four-dimensional effective action
   \bea
 S_{\rm eff} &=& \int d^4 x d^4 \theta\,  F(\Phi,W)= \tau_2(A)\, \int d^4 x \,{\rm tr}\, F^2+\ldots
 \eea
 where $\tau_2(A)$ is given by the second derivative of the prepotential.  Notice that despite the
  similarities the instantons contributing to  $\tau_2(A)$ are exotic and therefore the structure
  of $\tau_2(A)$ will be very different from that following from Seiberg-Witten type geometries.
  In particular, on the contrary of the prepotentials found in \cite{Seiberg:1994aj,Seiberg:1994rs}, the Casimir $A_{m}$'s appear
  in $\tau_2(A)$ only in a polynomial form.

The first terms in the expansion of the prepotential in the instanton winding number are given by
\bea &&
F(a,\epsilon) = 2\sqrt{ A_n}  \left[ q+\ft{1}{3} \,q^3 \,(
A_{n-2}-A_{n-3} (\ft{7}{4}\,g_2+\ft{9}{4}\, \epsilon_1 \epsilon_2) \right.\nn\\
&& \left. +A_{n-4}(\ft{13}{32}\, g_4+\ft{49}{32}\,g_2^2+\ft{45}{16}\,g_2\epsilon_1\epsilon_2 ))+\cdots\right]
+\ft12\, q^2 \left[ A_{n-1}-A_{n-2} (\ft{1}{4}\,g_2+\ft{1}{4}\epsilon_1\epsilon_2)\right.\nn\\
&& \left. +A_{n-3}(\ft{1}{16}\,g_4+\ft{1}{16}\,g_2\epsilon_1\epsilon_2)\right]+\cdots\,\,,
 \eea
 where $g_2=\epsilon_1^2+\epsilon_2^2$ and $g_4=\epsilon_1^4+\epsilon_2^4$,
 The case of $SO(4)$ gauge group is of particular interest. In this case the theory is conformal 
 and the instanton parameter $q$ dimensionless. Calculation of up to
 $q^7$ terms suggests that the all orders exact prepotential is
 \bea
F_{SO(4)}(\phi,G) = {\rm Pf}\, \phi \log\,\frac{1+q}{1-q}+\,
(\ft{1}{4}\,tr\,\phi^2-\ft{1}{16}\,tr\,G^2+\ft{1}{8}\,{\rm Pf}\,G)\log \,(1-q^2)\,,\nn\\
 \label{prep4}\eea
where 
\bea
tr\,\phi^2=-2( a_1^2+a_2^2);\quad tr\,G^2=-2g_2;\quad {\rm Pf}\,G=\epsilon_1 \epsilon_2\,.
\eea

 \section{Chiral ring}

The techniques developed in the previous sections apply as well to the computation of the general chiral correlator
$ tr\,\phi^J$ in the gauge theory.  These correlators constitute the so called "chiral ring".
   In complete analogy with the 4 d ${\cal N}=2$ SYM \cite{Losev:2003py,Nekrasov:2003rj,Flume:2004rp},  the
   generating function  $ tr\,\exp (\lambda \phi) $ of the chiral correlators    $\langle tr\, \phi^J\rangle$ can be represented as
\bea
\langle tr\,e^{\lambda \phi}\rangle  =\langle tr\,e^{\lambda \phi} \rangle_{cl}+{1\over Z}
\sum_K \,q^K\, \int d^k\chi \,\sum_i  \, \prod_{\ell}(1-T_{\ell}^\lambda)\,
e^{\lambda \chi_i}{\cal Z}_K(\chi)\,,
\label{ofi}\eea
where the factors $(1-T_{\ell}^\lambda)$ with $T_{\ell}\equiv e^{  \epsilon_\ell} $ properly take
care of the volume factor\footnote{The domain of integration of  (\ref{ofi}) is the entire moduli space. The latter
includes also the space-time translational zero modes whose contribution is cancelled by $\prod_{\ell}(1-T_{\ell})$.}.
${\cal Z}_K$ is the integrand  in the instanton partition functions  (\ref{zz}) and
 \be
 Z=\sum_K q^K \,\int \,d^k\chi \,{\cal Z}_K(\chi)
 \ee
is the partition function.
Thus to compute a specific correlator $\langle tr\,\phi^J\rangle $ in the contour integral one makes an insertion
 \bea
 O_{J,K}(\{ \chi_I\} )=\sum_{I=1}^K \left[\chi_I^J-\sum_{i=1}^4(\chi_I+\epsilon_i)^J+
 \sum_{i<j}^4(\chi_I+\epsilon_i+\epsilon_j)^J\right. \nn\\
 \left.-\sum_{i<j<k}^4(\chi_I+\epsilon_i+\epsilon_j+\epsilon_k)^J+(\chi_I+\epsilon_1+
 \epsilon_2+\epsilon_3+\epsilon_4)^J\right]\,
 \eea
 so that
 \be
 \langle tr\,\phi^J\rangle={1\over Z} \sum_K \,q^K\, \int d^k\chi \,{\cal Z}_K(\chi)  \, O_{J,K}(\{ \chi \})
\ee
 Remarkably, the normalized correlators are, unlike the partition function $Z$ itself,
   well defined even at the limit when $\epsilon$'s vanish. Direct calculations up to $q^3$ are not difficult:
 \bea
&&\langle tr\,\phi^2\rangle =-2\sum_{u=1}^n a_u^2 \nn\\
&&\langle tr\,\phi^4\rangle =2\sum_{u=1}^n a_u^4+192 \sqrt{A_n}\,
\,q-96\,A_{n-2}\,q^2+768 \sqrt{A_n}\,A_{n-4}\,q^3+\cdots\nn\\
&&\langle tr\,\phi^6\rangle =-2\sum_{u=1}^n a_u^6+1440\,A_{n-1}\,q^2-7680\,\sqrt{A_n}\,A_{n-3}\,q^3+\cdots\nn\\
&&\langle tr\,\phi^8\rangle
=2\sum_{u=1}^n a_u^8-6720\,A_{n}\,q^2+35840\,
\sqrt{A_n}\,A_{n-2}\,q^3+\cdots  \eea
where the first term gives the classical contribution to the correlator.
To avoid lengthy
expressions we did not present the gravitational corrections here but they can be obtained in a similar way.

  The first non trivial correlator $\langle tr\,\phi^4\rangle$ in the list can be related to the derivative of the
  prepotential.
This can be seen using the identity
 \bea
O_{4,K} (\{ \chi_I \})=24 K\, \epsilon_1
\epsilon_2 \epsilon_3 \epsilon_4
\eea
that implies
\bea \langle tr\,\phi^4\rangle=2\sum_{u=1}^n a_u^4+24\, q\partial_q
F= tr\,\phi_{\rm cl}^4+24\, \sum_K K F_K q^K
   \eea
Alternatively this can be seen by noticing that  each $\phi\sim F_{\mu\nu} \theta \gamma^{\mu\nu}$
in   $ tr\,\phi^4 $ soaks precisely two out of the eight fermionic zero modes in the instanton
background   and therefore all together the amplitude in the K-instanton
sector is given by $\int t_8 tr\, F^4\sim K$ time the normalized centered partition function $F_K$ \cite{Fucito:1996ua}.
    The remaining correlators $tr\,\phi^J$ with $J>4$ give new informations about the chiral ring
    of the theory beyond the prepotential.

    In a similar way we can compute chiral correlators in the case of
$\R^6\times \R^4/\Z_2$. Now the relevant insertion is
 \bea
 O_J(\chi_I,K)=\sum_{I=1}^K \left[\chi_I^J-(\chi_I+\epsilon_1)^J-
 (\chi_I+\epsilon_2)^J+ (\chi_I+\epsilon_1+\epsilon_2)^J\right]
 \eea
and one finds
\bea
\langle tr\,\phi^2\rangle =&&-\sum_{u=1}^n2a_u^2 -4\sqrt{A_n}\,\,q -2
\,A_{n-1}\,q^2-4\sqrt{A_n}\,A_{n-2}\,q^3\nn\\&&-2(A_{n-2}A_{n-1}+5A_nA_{n-3})q^4\nn\\&&-
4\sqrt{A_n}\,(A_{n-2}^2+3A_{n-1}A_{n-3})\,q^5 +\cdots\nn\\
\langle tr\,\phi^4\rangle =&&\sum_{u=1}^n2a_u^4-12A_n \,q^2 -16\sqrt{A_n}\,A_{n-1}\,q^3 \nn\\
&&-6(A_{n-1}^2+6A_{n}A_{n-2})q^4-48(A_{n-1}A_{n-2}+A_nA_{n-3})q^5+\cdots\nn\\
\langle tr\,\phi^6\rangle =&&-\sum_{u=1}^n2a_u^6 -40\sqrt{A_n}\,A_{n}\,q^3-90A_nA_{n-1}q^4\nn \\
&&-72 \sqrt{A_n}\,(A_{n-1}^2+3A_nA_{n-2}\,)q^5 +\cdots\nn\\
\langle tr\,\phi^8\rangle =&&\sum_{u=1}^n2a_u^8-140 A_{n}^2 q^4 -448\sqrt{A_n}\,A_{n}\,A_{n-1}q^5+\cdots
\eea
In this case the Matone relation takes the form
\bea \langle tr\,\phi^2\rangle=-2\sum_{u=1}^na_u^2-2 q\partial_q
F\label{matone}\eea

Note also that calculations up to $q^7$ strongly suggest that in the conformal case $SO(4)$ the exact expression
for the $\langle tr\,\phi^2\rangle$ is
\bea
\langle tr\,\phi^2\rangle =-2(a_1^2+a_2^2-\ft14 \,q^2(\epsilon_1^2+
\epsilon_1\epsilon_2+\epsilon_2^2))\,\frac{1}{1-q^2} -4a_1a_2 \,\frac{q}{1-q^2}\,\,.
\eea
This result is easily derived from (\ref{prep4}) after taking the derivative with respect to $q$, according to (\ref{matone}).

\vskip 1cm \noindent {\large {\bf Acknowledgments}} \vskip 0.2cm
The authors would like to thank M. Bianchi, M. Bill\`o, M.L. Frau and A. Lerda for many interesting discussions.
This work was partially supported by
the European Commission FP7 Programme Marie Curie Grant Agreement PIIF-GA-2008-221571 and
the Advanced Grant n.226455, ``Supersymmetry, Quantum Gravity and Gauge Fields" (SUPERFIELDS)
and by the Italian MIUR-PRIN contract 20075ATT78.
\noindent \vskip 1cm

\begin{appendix}

\section{SO(2n) prepotential}
Here we give the expression for (\ref{prepon}) up to $5$ instantons
(i.e. up to $q^5$) and up to the $4$th order in the gravitational corrections
 \bea
&& F(a_u,\epsilon_\ell)= \nn \\ && 8 \sqrt{A_n} \,\, q+q^2(  -2
A_{n-2}+\ft14\, A_{n-3} g_2-\ft{1}{64}\,  A_{n-4} (g_2^2+2 g_4))\nn\\
&&+8\sqrt{A_n}\,\, q^3(\ft{4}{3} \, A_{n-4}-\ft56\, A_{n-5} g_2
+\ft{1}{96}\,
A_{n-6} (25 g_2^2+34 g_4))\nn\\
&&+q^4\left[-\ft{1}{2}\,A_{n-3}^2-A_{n-2}A_{n-4}-17A_{n-1}A_{n-5}-
113A_{n}A_{n-6}\right.\nn\\&&+
\ft{1}{8}\,(3A_{n-3}A_{n-4}+19A_{n-2}A_{n-5}+195
A_{n-1}A_{n-6}+1491 A_{n}A_{n-7} )g_2\nn\\ &&-
\ft{3}{128}\,A_{n-4}^2(g_2^2+2g_4)-\ft{1}{64}\,
A_{n-3}A_{n-5}(15g_2^2+14g_4)\nn\\&&-\ft{3}{64}\,A_{n-2}A_{n-6}(41g_2^2+50g_4)-
\ft{3}{64}\,A_{n-1}A_{n-7}(389g_2^2+442g_4)\nn\\
&&\left.-\ft{1}{64}\,A_nA_{n-8}(9515g_2^2+11414g_4) \right]\nn\\
&&+8\sqrt{A_n}\,\,q^5\left[\ft{4}{5}\, (\ft{3}{2}\,A_{n-4}^2+7A_{n-3}A_{n-5}+
23A_{n-2}A_{n-6}+87A_{n-1}A_{n-7}+263A_{n}A_{n-8})\right.\nn\\
&& +\, \ft{1}{2}\,(-9A_{n-4}A_{n-5}-29A_{n-3}A_{n-6}-
109A_{n-2}A_{n-7}-413 A_{n-1}A_{n-8}-1389 A_{n}A_{n-9})g_2 \nn\\
&&+\,\ft{1}{160}\,A_{n-5}^2(205g_2^2+218g_4)+\,\ft{3}{80}\,A_{n-4}A_{n-6}(145g_2^2+154g_4)\nn\\
&&+\,\ft{1}{80}\,A_{n-3}A_{n-7}(1625g_2^2+1738g_4)+\,\ft{1}{80}\,A_{n-2}A_{n-8}(6325g_2^2+6802g_4)\\
&&\left.
+\,\ft{1}{80}\,A_{n-1}A_{n-9}(24265g_2^2+26378g_4)+\,\ft{1}{80}\,A_{n}A_{n-10}(86645g_2^2+97522g_4)\right]+
\cdots
\label{prep2} \eea

 \section{SO(8) case}
\label{aso8}
The case $n=4$ is special in the sense that q is dimensionless and
the theory is conformal.
Putting $n=4$ in (\ref{prep2}) and promoting the
expectation values to the respective fields,
\bea
tr\,\phi^2=-2\sum_{u=1}^4 a_u^2\,; \quad tr\,\phi^4=2\sum_{u=1}^4
a_u^4\,; \nn\\ tr\,G^2=-2\sum_{\ell=1}^4 \epsilon_\ell^2\,;\quad
tr\,G^4=2\sum_{\ell=1}^4 \epsilon_\ell^4
\eea
one finds
\begin{eqnarray}
F(\phi, G)&=&8 {\rm Pf} \phi\, (q+\ft{4}{3}\,
q^3+\ft{6}{5}\,q^5+\cdots )\nn \\
&+&\ft{1}{2}\,tr\,\phi^4\,
(q^2+\ft{1}{2}\,q^4+\cdots)-\ft{1}{4}\,\left(tr\,\phi^2\right)^2
(q^2+q^4+\cdots )\nn \\
&+&\left(\ft{1}{16}\,tr\,\phi^2\,\,tr\,G^2-\ft{1}{64}\,tr\,G^4-\ft{1}{256}\,
(tr\,G^2)^2\right)(q^2+\ft{3}{2}\,q^4+\cdots)\nn
\end{eqnarray}
This suggests that the exact expression in all orders of $q$ would
be
\begin{eqnarray}
F(\phi, G)&=&\sum_{k=0}^\infty\left\{8 {\rm Pf}\phi\, \sum_{l|2k+1}
\frac{1}{l}\,\,q^{2k+1}\nonumber \right.\\
&+&\ft{1}{2}tr\phi^4 \sum_{l|k}
\ft{1}{l}\,\left(q^{2k}-q^{4k}\right)
-\ft{1}{4}\left(tr\,\phi^2\right)^2 \sum_{l|k}
\frac{1}{l}\,\left(q^{2k}-\ft{1}{2}\,q^{4k}\right)
\nn \\
&+&\left.\left(\ft{1}{16}\,tr\,\phi^2\,\,tr\,G^2-\ft{1}{64}\,tr\,G^4-\ft{1}{256}\,
\left(tr\,G^2 \right)^2\right)\sum_{l|k} \frac{1}{l}\,\,q^{2k}
\right\}
\end{eqnarray}
 in agreement with the heterotic
results \cite{Lerche:1988zy,Lerche:1987qk,Lerche:1998nx,Bachas:1997mc,Bachas:1996bp,Forger:1998ub,Gava:1999ky,Gutperle:1999xu}  (
  up to normalizations
 of fields and traces  )
 \bea
 \Delta_{F^4} &=& -\log{ |\eta(T)|^4\over |\eta({T\over 2})|^4} +\ldots \nn\\
  \Delta_{(F^2)^2 } &=& - \ft12\log{ T_2 U_2 |\eta(T/2)|^8 |\eta(U)|^4 \over |\eta(T)|^4} +\ldots \nn\\
 \Delta_{R^4} &=& 4  \Delta_{(R^2)^2}=2  \Delta_{R^2 F^2}=-\ft{1}{16}  \log{ T_2 U_2 |\eta(T/2)|^4 |\eta(U)|^8}  +\ldots
 \eea
 where dots refer to moduli independent contributions.
 \end{appendix}
\providecommand{\href}[2]{#2}\begingroup\raggedright\endgroup

\end{document}